\def\year{2019}\relax
\pdfoutput=1
\documentclass[letterpaper]{article} 
\usepackage{aaai19}  
\usepackage{times}  
\usepackage{helvet}  
\usepackage{courier}  
\usepackage{url}  
\usepackage{graphicx}  
\usepackage{color}

\newcommand{\presec}{\vspace{-0.06in}}
\newcommand{\postsec}{\vspace{-0.03in}}
\newcommand{\presub}{\vspace{-0.06in}}
\newcommand{\postsub}{\vspace{-0.03in}}
\newcommand{\preeq}{\vspace{-1.5pt}}
\newcommand{\posteq}{\vspace{-1.5pt}}

\usepackage{amsmath, amsfonts}
\usepackage{subcaption}

\usepackage{graphics}      
\usepackage[noend]{algorithmic}
\usepackage{algorithm}
\usepackage{color}
\graphicspath{{./Figures/}}

	\newcommand{\tabincell}[2]{\begin{tabular}{@{}#1@{}}#2\end{tabular}}

\begin{document}
%
\title{
Deep Reinforcement Learning for Green Security Games with Real-Time Information 
}
\author{Yufei Wang$^1$, Zheyuan Ryan Shi$^2$, Lantao Yu$^3$, Yi Wu$^4$, Rohit Singh$^5$, Lucas Joppa$^6$, Fei Fang$^2$ \\
$^1$Peking University, $^2$Carnegie Mellon University, $^3$Stanford University\\ $^4$University of California, Berkeley, $^5$World Wild Fund for Nature, $^6$Microsoft Research
}
\maketitle

%








\sloppy
\begin{abstract}

Green Security Games (GSGs) have been proposed and applied to optimize patrols conducted by law enforcement agencies in green security domains such as combating poaching, illegal logging and overfishing. However, real-time information such as footprints and agents' subsequent actions upon receiving the information, e.g., rangers following the footprints to chase the poacher, have been neglected in previous work. 
To fill the gap,
we first propose a new game model GSG-I
which augments GSGs with sequential movement and the vital element of real-time information. Second, we design a novel deep reinforcement learning-based algorithm, DeDOL,
to compute a patrolling strategy that adapts to the real-time information against a best-responding attacker. 
DeDOL is built upon the double oracle framework and the policy-space response oracle, solving a restricted game and iteratively adding best response strategies to it through training deep Q-networks.
Exploring the game structure, DeDOL uses 
domain-specific heuristic strategies
as initial strategies and constructs several local modes
for efficient and parallelized training. 
To our knowledge, this is the first attempt to use Deep Q-Learning for security games.
\end{abstract}
\presec
\section{Introduction}
\postsec
Security games~\cite{tambe2011security} have been used for addressing complex resource allocation and patrolling problems in security and sustainability domains, with successful applications in critical infrastructure protection, security inspection and traffic enforcement~\cite{basilico2009leader,durkota2015optimal,yin2014game,rosenfeld2017traffic}.
In particular, Green Security Games (GSG) have been proposed to model the strategic interaction between law enforcement agencies (referred to as defenders) and their opponents (referred to as attackers) in green security domains such as combating poaching, illegal logging and overfishing. Mathematical programming based algorithms are designed to compute the optimal defender strategy, which prescribes strategically randomized patrol routes for the defender \cite{fang2015security,fang2016deploying,xu2017optimal}. 

Despite the efforts, a key element, real-time information, which exists widely in practice in green security domains, has been neglected in previous game models, not to mention the agents' subsequent actions upon receiving the information. 
For example, rangers can
observe traces left by the poacher (e.g., footprints, tree marks) or learn of poacher's location in real time from camera traps and conservation drones. A well-trained ranger would make use of the real-time information to adjust her patrol route.
Indeed, stories have been reported that rangers arrested the poachers after finding blood stains on the ground nearby~\cite{MRT2011news}.
Similarly, a poacher may also observe the ranger's action in real time and adjust his attack plan, and the rangers should be aware of such risk. 
Thus, the prescribed patrol plans in previous work have limited applicability in practice as they are not adaptive to observations during the patrol.

Our paper aims at filling the gap.
First, we propose a new game model GSG-I which augments GSGs with the vital element of real-time information and allows players to adjust their movements based on the received real-time information. These features lead to significant complexity, inevitably resulting in a large extensive-form game (EFG) with imperfect information.
Second, we design a novel deep reinforcement learning (DRL)-based algorithm, DeDOL (Deep-Q Network based Double Oracle enhanced with Local modes), to compute a patrolling strategy that adapts to the real-time information for zero-sum GSG-I. 
DeDOL is among the first few attempts to leverage advances in DRL for security games~\cite{kamra2018policy,trejo2016adapting} and is the first to use deep Q-learning for complex extensive-form security games.
DeDOL builds upon the classic double oracle framework (DO)~\cite{mcmahan2003planning,bosansky2013double-oracle} which solves zero-sum games using incremental strategy generation, and a meta-method named policy-space response oracle (PSRO)~\cite{lanctot2017unified} which augments DO with RL to handle a long time horizon in multi-agent interaction. Tailored towards GSG-I, DeDOL uses a deep Q-network (DQN) to compactly represent a pure strategy, integrates several recent advances in deep RL to find an approximate best response, which is a key step in the DO framework.
Further, DeDOL uses 
domain-specific heuristic strategies, including a parameterized random walk strategy and a random sweeping strategy as initial strategies to warm up the strategy generation process. In addition, exploring the game structure of GSG-I, DeDOL uses several local modes, each corresponding to a specific entry point of the attacker, to reduce the complexity of the game environment for efficient and parallelized training.

Finally, we provide extensive experimental results to demonstrate the effectiveness of our algorithm in GSG-I. We show that the DQN representation in DeDOL is able to approximate the best response given a fixed opponent. In small problems, we show that DeDOL achieves comparable performance as existing approaches for EFGs such as counterfactual regret  (CFR) minimization. In large games where CFR becomes intractable, DeDOL can find much better defender strategies than other baseline strategies.

\presec
\section{Preliminaries and Related Work}
\postsec
\subsubsection{Stackelberg Security Games (SSG) and Green Security Games} 
GSGs are a special class of SSG~\cite{tambe2011security,pita2008deployed,fang2013optimal}. In a GSG~\cite{fang2015security,basak2016combining}, 
a defender and an attacker interact in an area discretized into a grid of targets. The defender strategically allocates a limited number of patrol resources to patrol routes. 
The attacker chooses a target to attack. 
Each target is associated with reward and penalty for the defender and the attacker, representing the payoff to them depending on whether the attack on the target is successful.
Existing literature on GSGs and SSGs widely employs the solution concept of Strong Stackelberg Equilibrium (SSE), where the defender commits to a strategy that maximizes her expected utility assuming the attacker observes her strategy and best responds to it. 

When the game is zero-sum, common solution concepts such as Nash equilibrium (NE), SSE, Minimax, and Maximin, coincide, and a DO framework~\cite{mcmahan2003planning} is commonly applied to solve the game efficiently when the action space is large. 
DO is an iterative algorithm where in each iteration an NE is computed for a restricted game, in which each player only has a subset of pure strategies. Each player then adds a best response strategy against the opponent's current NE strategy to the restricted game. DO terminates when each player's best response strategy is already included in the restricted game. DO is guaranteed to converge to an NE of the original two-player zero-sum game. 

Most of the literature on SSG have neglected real-time information, with only a few exceptions~\cite{zhang2014defending} that are not designed for green security domains.

\presub
\subsubsection{Extensive-form Games (EFG)} 
\begin{figure}[t]
\centering
\includegraphics[width=.4\textwidth]{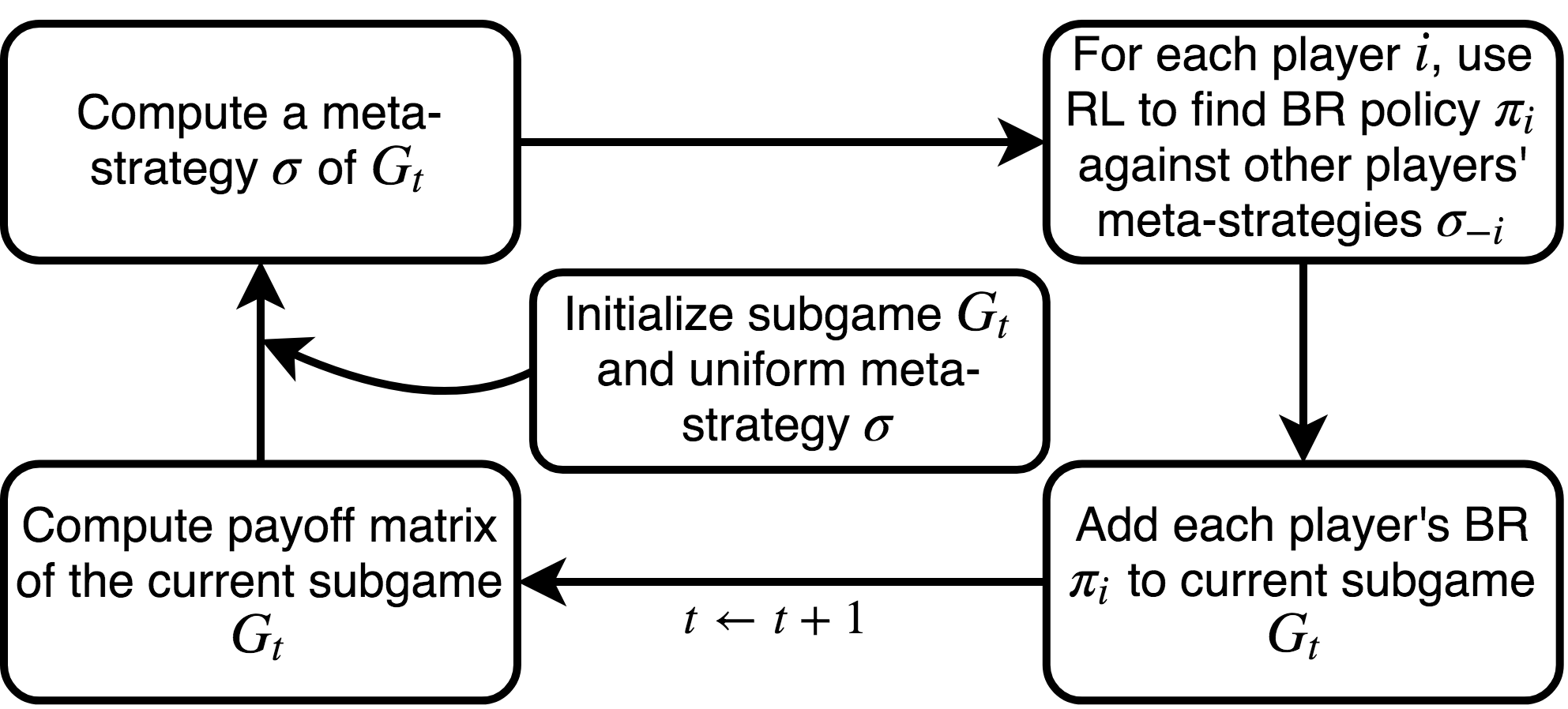}\vspace{-4pt}
\caption{The PSRO framework
}\label{fig:psro}
\vspace{-15pt}
\end{figure}

EFGs capture the sequential interaction between the players, and often presents more computational challenges than normal-form games~\cite{letchford2010computing}. Advanced algorithms for solving large-scale two-player zero-sum EFGs with imperfect information use counterfactual regret (CFR) minimization~\cite{zinkevich2008regret}, first-order methods~\cite{kroer2017smoothing},
abstraction~\cite{brown2017safe,vcermak2017algorithm},
or mathematical programming-based approach enhanced with the DO framework~\cite{bosansky2013double-oracle}.
Despite the huge success in solving poker games whose game tree is wide but shallow~\cite{brown2017libratus,moravvcik2017deepstack,bowling2015heads}, 
these approaches are not applicable to GSG-I, as its game tree is prohibitively deep in contrast to poker games. 
For example, CFR requires traversing the full game tree in each iteration and will run out of memory on the large instances of GSG-I.

\presub
\subsubsection{Deep RL and Multi-Agent RL} 

Deep RL has recently been widely used in complex sequential decision-making, in both single agent and multi-agent settings~\cite{oh2015action,leibo2017multi,foerster2016learning}. They have led to successful applications in Atari games~\cite{mnih2015human}, Go~\cite{silver2016mastering}, and continuous action control~\cite{mnih2016asynchronous}. 
An RL problem is usually formulated as a Markov Decision Process (MDP), comprising the state space $S$, action space $A$, transition probability $P$, reward function $r$, and the discounting factor $\gamma$. 
Q-learning~\cite{watkins1992q} is a popular value-based RL methods for discrete action space. The Q-value of a state-action pair $(s, a)$ under policy $\pi$ is defined as $Q^{\pi}(s_t, a_t) = \mathbb{E}_{s,a \sim \pi}[\sum_{l=0}^\infty \gamma^lr(s_{t+l}, a_{t+l}) | s_t, a_t]$.
DQN~\cite{mnih2015human} uses a deep neural network $Q^{\theta}$ to learn the optimal $Q$ value $Q^*(s,a) = max_{\pi}Q^{\pi}(s,a)$, by 
storing transitions $\{s,a,r,s'\}$ in an off-line replay buffer and minimizing the following loss: 
\begin{eqnarray}
\preeq
\small
\label{eq:DQN}
    \mathcal{L}(\theta) = \mathbb{E}_{s,a,r,s'}[(Q^{\theta}(s,a) - (r + \gamma \max_{a'} Q^{\tilde{\theta}}(s', a'))^2]
\posteq\posteq
\end{eqnarray} 
where $Q^{\tilde{\theta}}$ is the target Q network whose parameters $\tilde{\theta}$ are periodically copied from $\theta$ to stabilize training.
Besides Q-learning, Policy Gradient~\cite{sutton2000policy} is another kind of popular RL method. It employs a parametric stochastic policy $\pi_\theta$, and updates $\theta$ by gradient ascent according to the following theorem:
\begin{equation}
\label{eq::pg}
\preeq
\nabla_\theta E_{\pi_\theta}[r] = E_{s,a \sim \pi_\theta}[\nabla_\theta log\pi_\theta(a|s) \cdot Q^{\pi_\theta}(s,a)]
\end{equation}


A recent progress in multi-agent RL is the PSRO method~\cite{lanctot2017unified} (illustrated in Figure \ref{fig:psro})  
which generalizes DO by extending the pure strategies in the restricted game to parametrized policies and using deep RL to compute an approximate best response. PSRO provides a tractable approach for multi-player games with a long time horizon. However, since training in deep RL is time-consuming, it can only run a very limited number of iterations for large games, far less than needed for the convergence of DO. Thus, it may fail to find good strategies.
We propose several enhancements to PSRO to mitigate this concern, and provide a concrete implementation for GSG-I.



\presub
\subsubsection{Other Related Work} Patrolling game is an EFG where a patroller moves on a graph and an attacker chooses a node to ``penetrate''~\cite{agmon2008multi,basilico2009leader,horak2017heuristic}. Our game model extends patrolling games by allowing multiple attacks and partial observability of traces of movement in real time for both the defender and attacker.

\presec
\section{Green Security Game with Real-Time Information}
\postsec
\begin{figure}[t]
\centering
\includegraphics[width=0.78\linewidth]{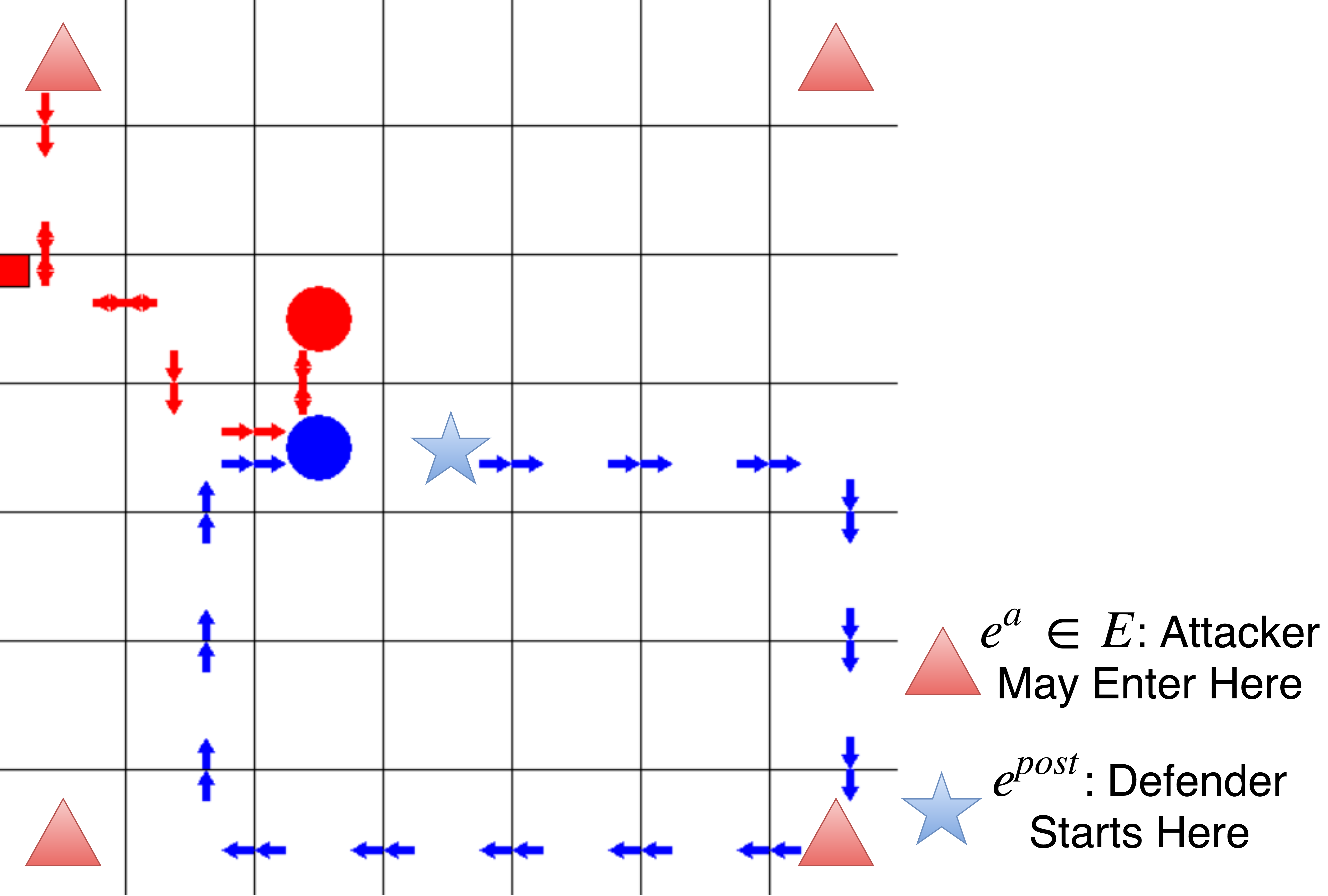}
\caption{Illustration of GSG-I. The red dot and blue dot represent the attacker and defender respectively. The red arrows and blue arrows represent their corresponding footprints. 
The red squares on the upper left corner of some cells represent the attack tools placed by the attackers. Each player only observes the opponent's footprint in their current cell.
}\label{fig:game}
\vspace{-14pt}
\end{figure}
In this section, we introduce GSG-I, Green Security Game with Real-Time Information.
As shown in Figure \ref{fig:game}, the basic environment of GSG-I is a grid world, with each cell representing a unique geographic area.
The game has two players, the defender and the attacker. 
At the beginning of the interaction, the attacker randomly chooses an entry point $e^a$ from a set $E$ of entry points,
while the defender always starts from the patrol post $e^{post}$. The attacker starts with a limited number of attack tools and uses them to set attacks at his desired locations. Such a game model is appropriate for a number of green security domains, e.g., rangers patrol in a conservation area to prevent poaching by removing animal snares and arresting poachers.
At each time step, the defender picks an action from her action space $\mathcal{A}^{d}$: \emph{\{up, down, right, left, stand still\}}. Simultaneously, the attacker picks an action from his action space $\mathcal{A}^{a}$: \emph{\{up, down, right, left, stand still\}} $\times$ \emph{\{place attack tool, not place attack tool\}}. 

Suppose an attack tool has been placed in a cell with coordinate $(i,j)$. At each time step, the attack tool successfully launches an attack with probability $P_{i,j}$. After a successful attack, the tool will be removed from the system. 
The defender tries to remove the attack tools prior to the attack and catch the attacker to stop him from placing more attack tools. She receives a positive reward $r^{tool}_{i,j}$ when she removes an attack tool from cell $(i,j)$, a positive reward $r^{catch}$ on catching the attacker, and a negative reward $p^{attack}_{i,j}$ when an attack tool launches an attack at cell $(i,j)$. 
The interaction ends either when the defender finds the attacker and all the attack tools, or 
when a maximum time step $T$ is reached. 
The defender's final payoff is the cumulative reward in the game. The attacker receives (positive or negative) rewards corresponding to these events as well and in this paper we focus on zero-sum games.

As shown in Figure \ref{fig:game}, both players leave footprints as they move around. There can be many other forms of real-time information such as dropped belongings and local witnesses, yet for simplicity we use only footprints in this paper. 
In our game we assume both players have only \emph{local observations}. They only observe their opponent's footprints in the current cell rather than the full grid, reflecting the fact that they often have a limited view of the environment due to the dense vegetation, complex terrain, or formidable weather. 
We assume the players have unlimited memory and can keep a record of the observations since the beginning of each interaction. 

Hence, we define a player's pure strategy or policy in this game (we use policy and strategy interchangeably in this paper) as a deterministic mapping from his observation and action history to his action space. A player can employ a mixed policy, which is a probability distribution over the pure strategies. 
\presec
\section{Computing Optimal Patrol Strategy}
\postsec
It is nontrivial to find an optimal patrol strategy. Simple action rules such as following the footprints or escaping from the footprints may not be the best strategy as shown in experiments.
We now introduce DeDOL, our algorithm designed for computing the optimal defender's patrol strategy in zero-sum GSG-I. DeDOL builds upon the PSRO framework. Thus we will first introduce a DQN-based oracle for computing an approximate best response, and then introduce DeDOL, which uses the best response oracle as a subroutine.
\presub\presec
\vspace{-5pt}\postsub
\subsection{Approximating Player's Best Response}
We first consider an easier scenario where either player is static, i.e. using a fixed and possibly randomized strategy. The player's fixed strategy and the game dynamics of GSG-I then defines an MDP for the other player.
We represent the other player's policy by a neural network, and use reinforcement learning to find an empirically best response strategy. In this subsection we assume the defender is the learning player, as the method for the attacker is identical.

Due to the strong spatial patterns of GSG-I, we employ a convolutional neural network (CNN) to represent the defender policy $\pi_d(a^d_t | s^d_t)$, which is a mapping from her state space to her action space. The input to the CNN is the defender's state $s^d_t$, represented by a 3-D tensor with the same width and height as the grid world and each channel encoding different features. Specifically, the first 8 channels are the binary encodings of the local attacker footprints (\{four directions\} $\times$ \{entering or leaving\}); the next 8 channels are similar encodings of the defender's own footprints;
the 17th channel is one-of-K encoding that indicates the defender's current location; the 18th channel is the success probability of the attack tools of the grid world; the 19th channel is the normalized time step, which is the same for all cells.

\begin{figure}[t]
\centering
\includegraphics[width=.48\textwidth]{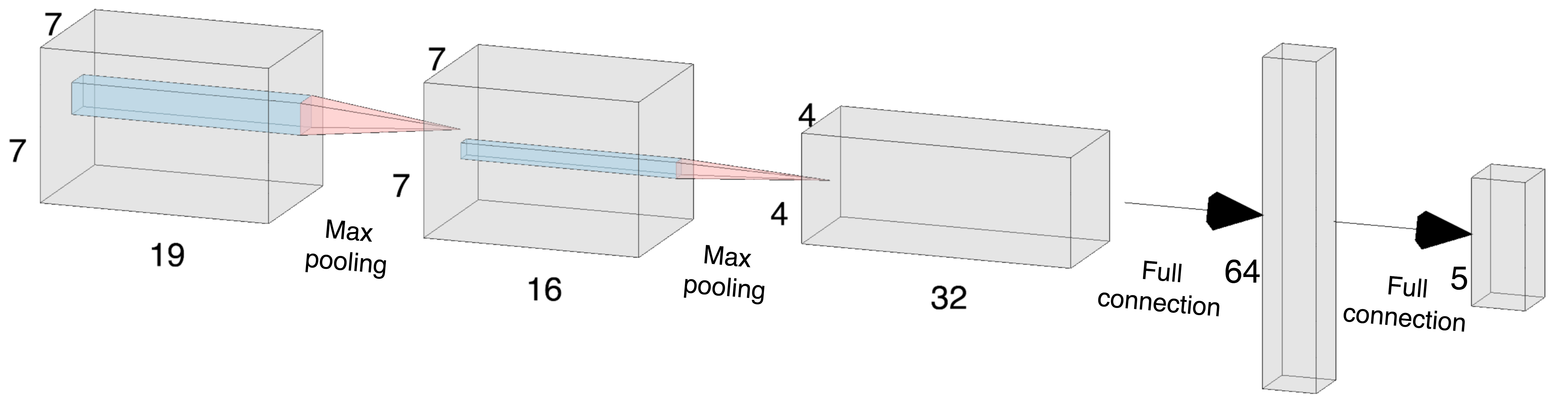}\vspace{-5pt}
\caption{The defender's neural network architecture for the $7 \times 7$ grid.
}\label{fig:nn}
\vspace{-15pt}
\end{figure}

Figure~\ref{fig:nn} shows the neural network architecture when the game has a $7 \times 7$ grid (the network architecture of other grid sizes is detailed in Appendix A). The first hidden layer is a convolutional layer with 16 filters of size $4 \times 4$ and strides $1 \times 1$. The second layer is a convolutional layer with 32 filters of size $2 \times 2$ and strides $2 \times 2$. Each hidden layer is followed by a $relu$ non-linear transformation and a max-pooling layer. The output layer is a fully-connected layer which transforms the hidden representation of the state to the final policy. 
For the DQN method, each output dimension represents to the Q-value of each action, and the neural network corresponds to a pure defender strategy where she takes the action with the highest Q-value. For the policy gradient method, the neural network corresponds to a stochastic policy where each output dimension represents the probability of choosing each action.

We use Deep Q-learning~\cite{mnih2015human} or Policy Gradient~\cite{sutton2000policy} to approximate the best response with the above neural network.
Due to the highly dynamic environment of GSG-I, the training of the vanilla versions of these two algorithms proved difficult, especially when the other player uses a randomized strategy.
Therefore, we employ the double DQN methods~\cite{van2016deep} to improve the stability of training, 
and the loss we minimize changes to:
\begin{equation*}
\preeq
\small
\mathcal{L}(\theta) = \mathbb{E}_{s,a,r,s'}[(Q^{\theta}(s,a) - (r + \gamma  Q^{\tilde{\theta}}(s', \arg\max_{a'} Q^{\theta}(s', a')))^2]
\posteq
\end{equation*}
Furthermore, we incorporate the dueling network architecture \cite{wang2016dueling} upon double DQN for more efficient learning.
For Policy Gradient methods, we implement the actor-critic algorithm \cite{konda2000actor}, where $Q^{\pi_\theta}(s,a)$ in Eq.\ref{eq::pg} is replaced by $r + \gamma V^{\pi_\theta}(s) - V^{\pi_\theta}(s')$ to lower the variance. We implement another CNN to approximate the state-value $V^{\pi_\theta}(s)$ by changing the last output layer in Figure \ref{fig:nn} to be a scalar.  
At last, we employ gradient clipping~\cite{clipping2013on} in the training of both methods to deal with the gradient exploding issue.

The reader might notice that this neural network-based representation does not capture all defender strategies. However, the strong expressiveness makes it a memory-efficient alternative. Furthermore, we show later that we lose little by using this compact representation. 

\presub
\subsection{The DeDOL Algorithm} 
\postsub

Having deep RL as the players' best response oracle lays the groundwork for finding the 
optimal defender strategy in GSG-I. In zero-sum GSG-I, the SSE strategy is also the NE strategy. 
The PSRO framework~\cite{lanctot2017unified} (Figure~\ref{fig:psro}) can be applied to compute the NE strategy. A naive implementation of PSRO in GSG-I is as follows: we use a randomly initialized DQN as the initial strategy for each player. At each iteration, we first get the payoff matrix for the current strategies by simulation and compute the NE for the current game matrix. Then we fix the NE strategy for one player and calculate the best response strategy of another player with DQN, as detailed above. 
We add these best response strategies of each player to the current game if they are better than the existing strategies, and repeat the procedure until no better responses can be found for either player.

However, as we will show in the experiment section, this naive implementation (referred to as Vanilla-PSRO throughout) does not perform well in practice due to the following limitations: 1) randomly initialized DQNs are rarely meaningful policies, and it takes several iterations for Vanilla-PSRO to evolve a reasonable set of strategies out of them. This problem is especially prominent as the training of DQN takes a nontrivial amount of time. 2) The computed best response in Vanilla-PSRO tends to overfit to the specific NE strategy that it was trained against, and may not be robust against other opponent strategies in the complex GSG-I. 3) In GSG-I, the attacker could enter the grid world through multiple entry points. Using a single best response defender DQN to deal with all these possibilities makes the training rather difficult and the learned strategies sub-optimal.   

Therefore, we propose DeDOL, which enhances Vanilla-PSRO by introducing three key elements as discussed below.
\presub
\subsubsection{Initial Strategies for DO}

The problem with naive initial strategies is that the best response against a highly exploitable strategy could still be highly exploitable itself. Thus, adding such a best response strategy to the strategy profile helps little. To alleviate this problem, we propose two lightweight yet effective domain-specific heuristic strategies as the initial strategies of DO.

For the attacker, we use a parameterized random walk policy. 
Suppose the current coordinate of the attacker is $(m,n)$ and the maximum coordinate on the map is $(M,N)$. We can define the average success probability for the \textit{up} direction as $\frac{1}{(m-1)\cdot N} \sum_{0 \leq i<m, 0 \leq j \leq N} P_{i,j}$. Recall that $P_{i,j}$ is the probability that an attack tool launches an attack successfully at cell $(i,j)$. Similarly, we can define the average success probability for all the other directions. For simplicity, we use an integer $k \in \{1,\ldots,5\}$ to denote one of the five directions (the fifth ``direction'' is for the action ``stand still''). This way, we can get an average success probability vector $\bar P \in \mathbb{R^+}^5$ ($\bar P_5$ is the success probability of the current grid). 
Another important factor that should be taken into consideration is the observed footprints. We use vectors $I \in \{0,1\}^5$ and $O \in \{0,1\}^5$ to represent the footprints states, where each dimension $I_k$ (or $O_k$) is a binary variable, indicating whether or not there is an entering (or leaving) footprint from that direction (for the fifth ``stand still'' direction, $I_5 = O_5 = 0$). Now we can define the parameterized heuristic policy for the attacker's movement as
\begin{equation}
\small
\pi_{a}(a^{a}_t = k|s^{a}_t) = \frac{\exp(w_p \cdot \bar P_k + w_i \cdot I_k + w_o \cdot O_k)}{\sum_{z} \exp(w_p \cdot \bar P_z + w_i \cdot I_z + w_o \cdot O_z)}
\end{equation}
where $w_p$, $w_i$ and $w_o$ are parameters for the average success probability, entering and leaving footprints, respectively. 

The success probability of the attack tool directly impacts the decision of where to place it. We define the probability of placing an attack tool in cell $(m,n)$ as
\begin{equation}
\small
\eta_{a}(b^{a}_t = 1|s^{a}_t) = \frac{\exp(P_{m,n} / \tau)}{\sum_i\sum_j \exp(P_{i,j} / \tau)}
\end{equation}
where $\tau$ is a temperature parameter.

The behavioral model as described above is boundedly rational. Real-world applications often feature bounded rationality due to various constraints. We use this parameterized heuristic policy as the initial attacker strategy in DeDOL with parameters set following advice from domain experts\footnote{We showed the experts the attacker behaviors with different parameter combinations using our designed GUI, and pick the one they think most reasonable.}.

For the defender's initial strategy, we could use a similar, and even simpler parameterized heuristic random walk policy, as the defender's decision only involves movement. However, here we propose to use another more effective policy, called random sweeping. In the beginning, the defender randomly chooses a direction to move in and heads towards the boundary. She then travels along the boundary until she finds any footprint from the attacker and follows the footprints. If there are multiple footprints at the same location, she randomly chooses one to follow. This turns out to be a very strong strategy, as to defeat this strategy, the attacker has to confuse the defender using his footprints.


\presub
\subsubsection{Exploration and Termination}
The best response against the NE of a subgame $G_t$, $Nash(G_t)$, may not generalize well against other unexplored strategies in a complex game like GSG-I. In DeDOL, the fixed player instead uses a mixture of $Nash(G_t)$ and $Unif(G_t)$, the uniform random strategy where the player chooses each strategy in $G_t$ with equal probability. That is, with probability $1-\alpha$ he plays $Nash(G_t)$, and with probability $\alpha$ he plays $Unif(G_t)$. 

As a result, the trained DQN is an (approximate) best response to the NE strategy mixed with exploration, rather than the NE itself. Therefore we need to check in subroutine VALID (Algorithm~\ref{algo:valid}) whether it is still a better response to the NE strategy than the existing strategies. This method is similar to the better response oracle introduced in~\cite{jain2013security}. If neither of the new strategies for the two players is a better response, we discard them and train against the NE strategies without exploration. The parent procedure Algorithm~\ref{algo:dedols} 
terminates if we again find no better  responses. Algorithm~\ref{algo:dedols} may also terminate if it is intended to run a fixed number of iterations or cut short by the user. Upon termination, we pick the defender NE strategy (possibly plus exploration) and the attacker's best response which together give the highest defender's expected utility.
\setlength{\textfloatsep}{5pt}
\begin{algorithm}[t]
	\caption{DeDOL-S
    }                       
	\begin{algorithmic}[1]  \label{algo:dedols}
		\REQUIRE Mode (local/global), attacker entry point (if local), initial subgame $G_0$, exploration rate $\alpha$
		\FOR{iteration $t$}
		\STATE Run simulations to obtain current game matrix $G_t$. 
        \STATE $Nash(G_t) = (\sigma^{d}_t, \sigma^{a}_t)$, $Unif(G_t) = (\rho^{d}_t, \rho^{a}_t)$.
		\STATE Train defender DQN $f^{d}_t$ against $(1-\alpha)\sigma^{a}_t + \alpha \rho^{a}_t$.
		\STATE Train attacker DQN $f^{a}_t$ against $(1-\alpha)\sigma^{d}_t + \alpha \rho^{d}_t$.
		\STATE VALID($f^{d}_t, f^{a}_t, G_t$)
		\IF{TERMINATE condition satisfied}
        \STATE $k^* = \arg\max_k \{defEU((1-\alpha)\sigma^{d}_{k} + \alpha \rho^{d}_k, f^{a}_k)$, and $defEU(\sigma^{d}_{k}, \overline{f^{a}_k})$ if any were ever calculated$\}$
		\RETURN Defender optimal strategy from the $k^*$th iteration per above, current subgame $G_t$
		\ENDIF
		\ENDFOR
	\end{algorithmic} 
\end{algorithm}
\begin{algorithm}[t]
	\caption{VALID}                       
	\begin{algorithmic}[1]  \label{algo:valid}
		\REQUIRE DQNs $f^{d}_t, f^{a}_t$, subgame $G_t$ with NE $(\sigma^{d}_t, \sigma^{a}_t)$ 
		\IF{$\sigma^{a}_t \cdot G_t(f^{a}, f^{d}_t) \geq \sigma^{a}_t \cdot G_t(f^{a}, f^{d}_k), \forall k < t$} \label{algoline:checkstart}
		\STATE Defender best response $f^{d}_t$ is valid, add to $G_t$
		\ENDIF
		\IF{$\sigma^{d}_t \cdot G_t(f^{d}, f^{a}_t) \geq \sigma^{d}_t \cdot G_t(f^{d}, f^{a}_k), \forall k < t$}
		\STATE Attacker best response $f^{a}_t$ is valid, add to $G_t$
		\ENDIF \label{algoline:checkend}
		\IF{neither of the above is true}
		\STATE Fix $\sigma^{a}_t$ from $G_t$, train defender DQN $\overline{f^{d}_t}$ against it.
		\STATE Fix $\sigma^{d}_t$ from $G_t$, train attacker DQN $\overline{f^{a}_t}$ against it.
		\STATE Do Lines~\ref{algoline:checkstart}-\ref{algoline:checkend} with $f^{d}_t$, $f^{a}_t$ replaced by $\overline{f^{d}_t}$, $\overline{f^{a}_t}$.
		\STATE If neither `if' is true again, signal TERMINATE
		\ENDIF
	\end{algorithmic} 
\end{algorithm}

\presub
\subsubsection{Local Modes}
\label{subsec::DQN_training}

\begin{figure}[t]
\centering
\includegraphics[width=0.9\linewidth]{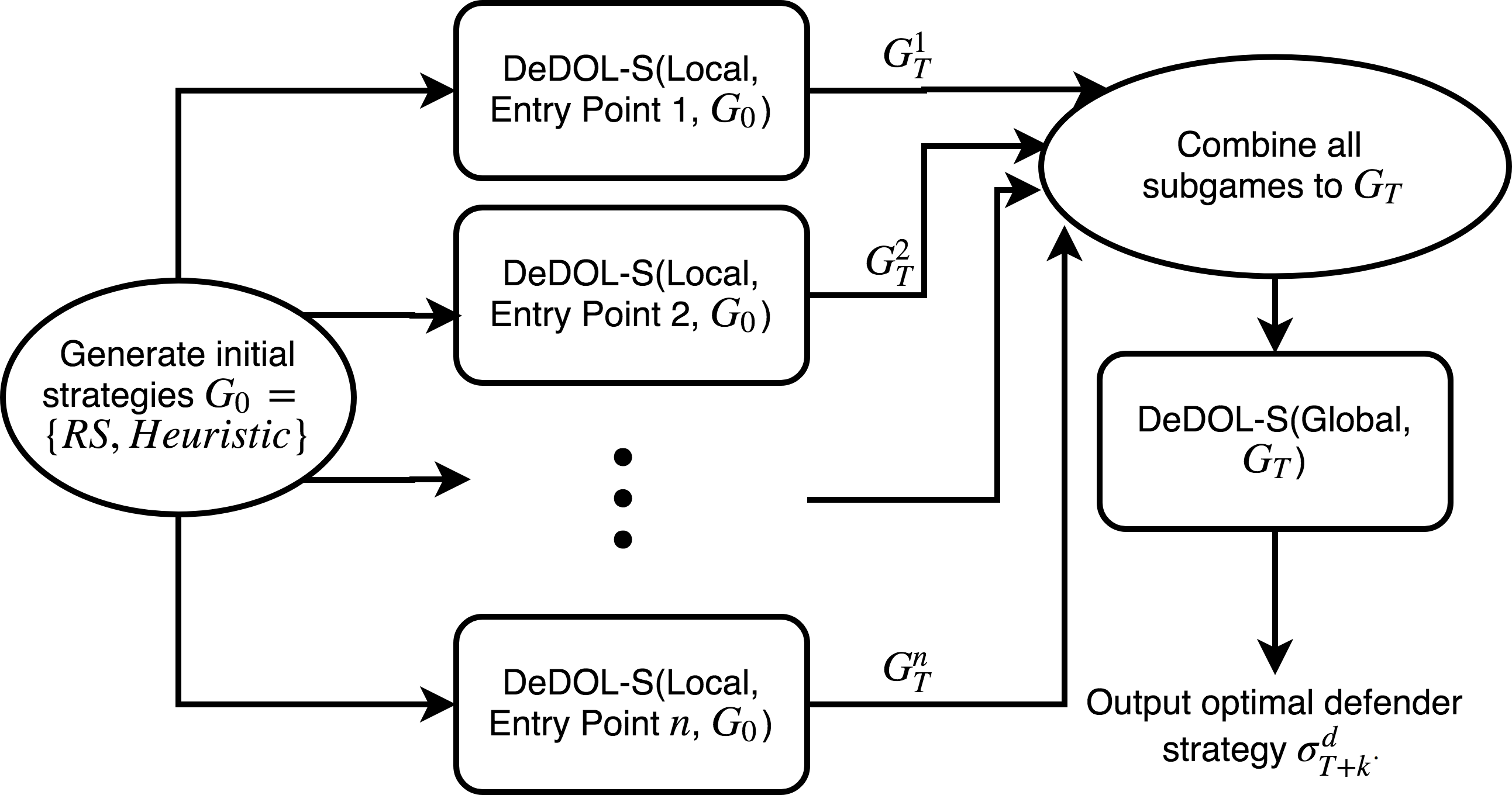}
\vspace{-5pt}
\caption{The DeDOL Algorithm}
\label{fig:DeDOL}
\end{figure}

We refer to the algorithm introduced so far as DeDOL-S (Algorithm~\ref{algo:dedols}). Our main algorithm DeDOL, illustrated in Figure~\ref{fig:DeDOL}, uses DeDOL-S as a subroutine. We now conclude this section by introducing the key feature of DeDOL: local modes.


Since it is challenging for a single defender DQN to approximate best response against an attacker entering from different cells, 
we divide the original game (referred to as the global mode) into several local modes. In each mode the attacker has a fixed entry location. In DeDOL, we first run DeDOL-S in each of the local modes in parallel. After a few iterations, we combine the DQNs trained in all local modes to form a new subgame. Then, we use this new subgame as the initial subgame and run DeDOL-S in the global mode for more iterations. 

When the attacker enters from the same location, both players (especially the defender) will face a more stable environment and thus are able to learn better strategies more quickly. More importantly, these strategies serve as good building blocks for the equilibrium meta-strategy in the global mode, thus improving the strategy quality. In the following section, we show that this approach performs better than several other variants.

\presec
\section{Experiments}
\postsec

\begin{figure}
\vspace{-0.02in}
\centering
\includegraphics[scale = 0.35]{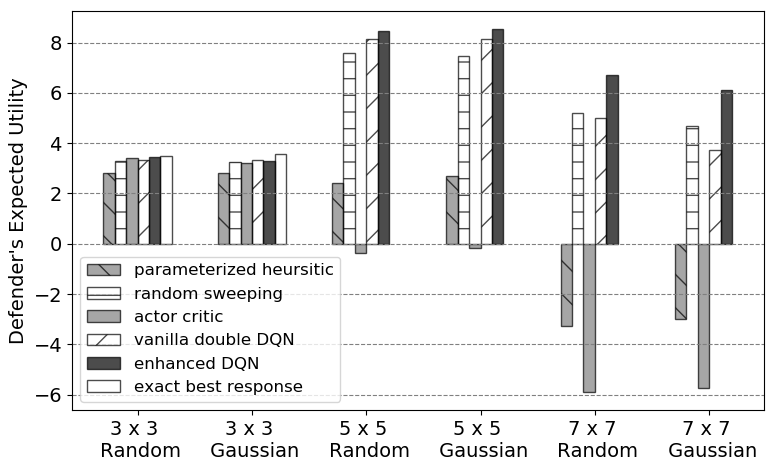}\vspace{-3pt}
\caption{Expected utilities of different patroller strategies against a parameterized random walk poacher.}
\label{fig:best_utility}
\vspace{-3pt}
\end{figure}

We test DeDOL in GSG-I using a case study on wildlife anti-poaching, where the grid world represents a wildlife conservation area. The attacker corresponds to a poacher carrying attack tools, i.e., snares, to catch animals. The defender corresponds to a patroller moving in the area to stop poaching by removing snares and arresting poachers. Each cell of the grid world has a corresponding animal density, which is proportional to the probability that a snare successfully catches an animal in that cell.
The animal densities are generated either uniformly randomly, or following a mixture Gaussian. The latter reflects that in reality the animal density is often higher along mountain ranges and decreases as we move away. The game environment of different types and sizes are shown in Appendix C.
We test DeDOL on three grid worlds of different sizes: $3 \times 3, 5\times 5$, and $7 \times 7$.
All experiments are carried out on Microsoft Azure standard NC6 virtual machines,  with a 6-core 2.60 GHz Intel Xeon E5-2690 CPU, a Tesla K80 GPU, and a 56G RAM.

\presub
\subsection{Best Response Approximation}
\postsub

\begin{figure}[t]
\centering
 \includegraphics[width=.4\textwidth]{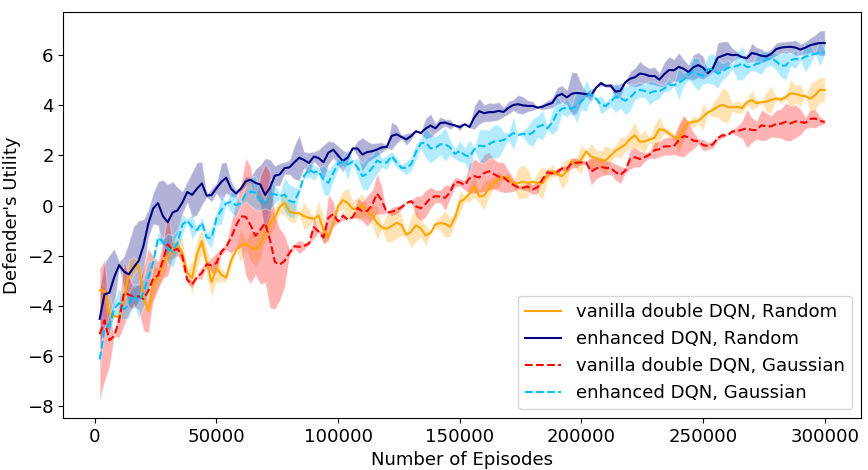}  
 \vspace{-6pt}
\caption{The learning curves of patroller DQNs against parameterized heuristic random walk poachers on $7 \times 7$ grids, averaged across
four runs.}
\label{fig:DQNlearning}
\end{figure}

The ability to approximate a best response strategy against a fixed opponent is foundational to our algorithm. Therefore, we start by comparing the performance of several methods against a parameterized heuristic poacher with parameters  set following the advice from domain experts. 
We compare the random sweeping strategy, parameterized random walk patroller with parameters set by grid search, the vanilla double DQN, the dueling double DQN + gradient clipping (enhanced DQN), and the actor-critic algorithm.
On the $7\times 7$ grid world, we train both DQNs and actor-critic using Adam optimizer~\cite{kingma2015adam:} with a learning rate of $0.0001$ for $300000$ episodes. 
More detailed training parameters are provided in Appendix B. Figure \ref{fig:DQNlearning} shows the learning curves of both DQNs in $7 \times 7$ grid. The actor-critic algorithm does not converge in our experiments.

In the smaller $3 \times 3$ game with 4 time steps, we can compute the exact poacher best response given a patroller strategy~\cite{bosansky2013double-oracle} (details in Appendix F).
However, this method becomes intractable with just a $5 \times 5$ grid which has 25 time steps and over $10^{20}$ information sets. 

\begin{figure*}[t]
\centering
\begin{tabular}{cccccc}
\includegraphics[width=.2\textwidth]{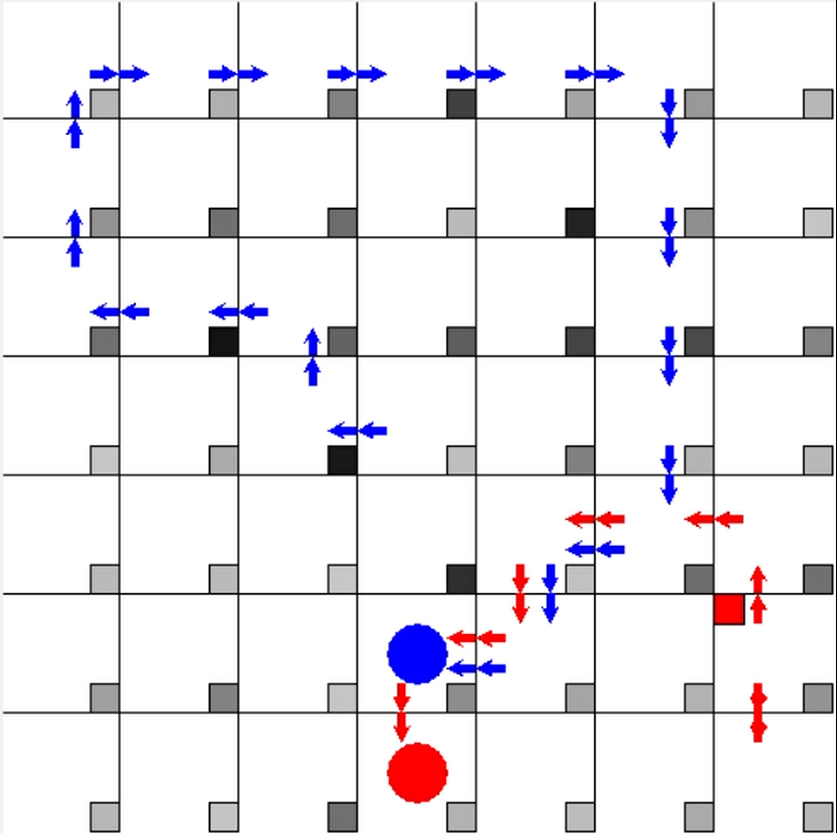}\vspace{-3pt}
& \quad &
\includegraphics[width=.2\textwidth]{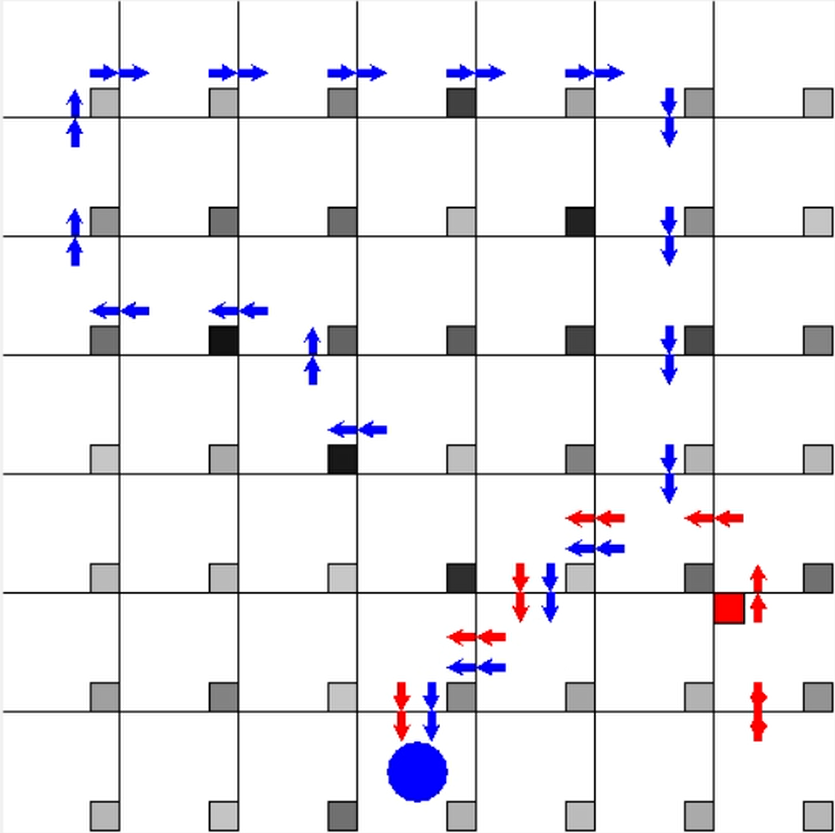}\vspace{-3pt}
& \quad &
\includegraphics[width=.2\textwidth]{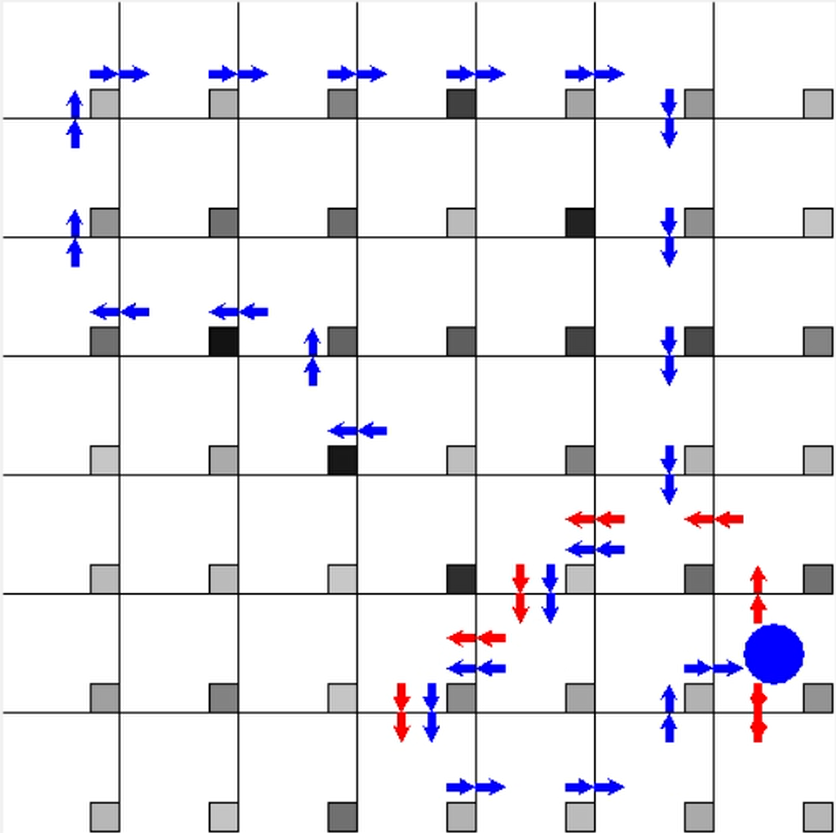}\vspace{-3pt}
\\
& \\
\vspace{-5pt}
(a) At time step 18 & $~~$\qquad\quad & (b) At time step 19 & \qquad $~~$\quad & (c) At time step 23 
\end{tabular}
\vspace{-2pt}
\caption{The learned patroller DQN strategy against a parameterized heuristic random walk poacher. Here, the darkness of the square in each cell indicates the animal density. }
\label{fig:paDQNstrategy}
\vspace{-4pt}
\end{figure*}

The results of each method are summarized in Figure \ref{fig:best_utility}. The enhanced DQN patroller achieves the highest expected utility among all compared strategies in all settings. Compared to the exact solution in the $3 \times 3$ game, the enhanced DQN is indeed a very good best response approximation.
Figure \ref{fig:paDQNstrategy} provides an illustration of the learned enhanced DQN strategy on a $7\times7$ grid with random animal density. Note that the enhanced DQN patroller cleverly learns to first patrol towards the corner on a path of high animal densities. She then moves along the boundary, and upon finding the poacher's footprints, follows them to catch the poacher. After the poacher is caught, she induces from the observed footprints that the entry point of the poacher should be the bottom right corner. Hence she patrols that area and successfully removes a snare there. 
A similar visualization for the trained poacher DQN against a random sweeping patroller is shown in Appendix D. We dropped the actor-critic algorithm in subsequent experiments as it performs poorly.


\presub
\subsection{Small Games}
\postsub

\begin{table}[t]
\vspace{-10pt}
\footnotesize
\centering
\begin{tabular}{c|c|c|c|c|c}
$\alpha$ &0 & 0.1 & 0.15 & 0.25 & 0.4 \\ \hline
$3 \times 3$ Random &  0.60 & 0.43 & \textbf{0.73} & 0.73 & 0.44 \\ \hline
$3 \times 3$ Gaussian & 0.12 & 0.39  & \textbf{0.64} & 0.37 & 0.04 \\ 
\end{tabular}
\caption{Patroller's expected utility with different exploration rate $\alpha$ on the $3 \times 3$ grid using DeDOL (global only).}
\label{tab:exploration rate}
\end{table}
 
\begin{table*}[t] 
\centering
\footnotesize
\begin{tabular}{c|c|c|c|c|c|c}
 & \tabincell{c}{ Random \\ Sweeping } &
\tabincell{c}{ Vanilla \\ PSRO} & 
\tabincell{c}{ DeDOL \\ Pure Global Mode} &
\tabincell{c}{ DeDOL \\ Local + Global Mode} & \tabincell{c}{ DeDOL \\ Pure Local Mode} &
CFR \\ \hline
$3\times3$ Random & -0.04 & 0.65 (16) & 0.73 (16) & \textbf{0.85} (10 + 2)& 0.71 (20)& \textbf{1.01} (3500) \\ \hline 
$3\times3$ Gaussian & -0.09 & 0.52 (16) & 0.75 (16) & \textbf{0.86} (10 + 2) & 0.75(20) & \textbf{1.05} (3500) \\ \hline
$5 \times 5$ Random & -1.91 & -8.98 (4) & -1.63 (4) & -0.42 (4 + 1) & \textbf{-0.25} (5) & - \\ \hline
$5 \times 5$ Gaussian &  -1.16 & -9.09 (4) & -0.43 (4)  & \textbf{0.60} (4 + 1) & -2.41 (5) & - \\ \hline 
$7 \times 7$ Random & -4.06 & -10.65 (4) & -2.00 (4) &  \textbf{-0.54} (3 + 1) & -1.72(5)&  - \\ \hline 
$7 \times 7$ Gaussian & -4.25 & -10.08 (4) & -4.15 (4) & \textbf{-2.35} (3 + 1) & -2.62(5) &-
\end{tabular}
\vspace{-4pt}
\caption{The highest patroller's expected utility among all DO / CFR iterations. The numbers in the parentheses show the finished DO / CFR iterations within the given running time. The highest value among all algorithms are in bold. The detail values of the defender expected utility at each iteration of DeDOL are shown in Appendix E.}
\label{tab:main_result}
\vspace{-13pt}
\end{table*}

Now we have shown that (enhanced) DQN can approximate a best response well, we move on to test the whole DeDOL algorithm. 
We first test it on a 3 $\times$ 3 grid. The game has $4$ time steps, and the attacker has $3$ snares. The full game tree has roughly $4.5 \times 10^7$ nodes. 

Before going into the main results, we tune the exploration rate $\alpha$ by running 16 iterations in DeDOL-S global mode. Table \ref{tab:exploration rate} shows the highest patroller's expected utility against an exact best response poacher with different exploration rate. Since $\alpha = 0.15$ achieves the best result in both map types, we set $\alpha$ to be 0.15 in the following experiments. The defender's utility with $\alpha = 0.15$ is also much higher than with $\alpha = 0$, showing that exploration is helpful.

We now compare the performance of DeDOL with other baselines in GSG-I.
To investigate whether DQNs trained in local modes would indeed help in global mode, we implement three versions of the DeDOL algorithm by controlling the number of DeDOL-S iterations in the local and global mode: 1) we run zero iteration in local modes, i.e., run DeDOL-S directly in global mode; 2) we run DeDOL-S in local modes for several iterations, return to the global mode and run for several more iterations (Fig.~\ref{fig:DeDOL});  3) we run DeDOL-S purely in the local modes, and upon termination, return to the global mode, compute an NE strategy and running no more iterations.



We use the counterfactual regret (CFR) minimization~\cite{zinkevich2008regret}, random sweeping, and Vanilla-PSRO as three baselines. Each learning algorithm runs for a day. In particular, in the local + global version of DeDOL, half a day is used for local modes and the rest for the global mode. With chance sampling, the CFR algorithm traverses roughly $4.3 \times 10^6$ nodes in one iteration, and finishes 3500 iterations in a day on our hardware.



The first two rows of Table \ref{tab:main_result} report the highest patroller's expected utilities, calculated against an exact best response poacher.
We note that the highest patroller's utility achieved by DeDOL is slightly lower than that of CFR given the same amount of time. However, DeDOL needs much less memory as it only needs to store several neural networks, while the CFR algorithm has to store the whole huge game tree.
The table also shows that all implementation versions of DeDOL outperform Vanilla-PSRO, and have much higher utility than the random sweeping baseline. In addition, the local + global modes version achieves the best result in both map types, which proves its effectiveness. Note that the local mode implementation finishes more iterations because the training of DQNs converges faster in its simpler environment.

\presub
\vspace{-0.05in}
\subsection{Large Games}
\postsub
We also perform tests on large games with $5 \times 5$ and $7 \times 7$ grid. $5 \times 5$ game has 25 time steps, and $7 \times 7$ game has 75 time steps. In both games, the attacker has 6 snares. 

We still implement 3 versions of DeDOL as detailed in the previous subsection. The running time on $5 \times 5$ grid is 3 days, and 5 days on the $7\times 7$ game. For the local + global version of DeDOL, we allocate 2 days for local mode on $5 \times 5$, and 3 days on $7 \times 7$. We report the performance of DeDOL in Table~\ref{tab:main_result}. As aforementioned, with even a $5 \times 5$ grid, there are over $10^{50}$ game states and $10^{20}$ information sets. Thus, CFR becomes intractable in terms of running time and memory usage, so is computing the exact best response. Therefore, in Table~\ref{tab:main_result} the patroller's expected utilities are calculated against their respective best response DQN poacher\footnote{Here, we train a separate DQN for a longer time than in a DO iteration. We also test against the poacher's heuristic strategy and pick the better one, which is always the DQN in the experiments.}.


Similar to the results in small games, all versions of DeDOL significantly outperform the Vanilla-PSRO and the random sweeping baseline. The Vanilla-PSRO performs extremely poor here because it starts with a poor randomly initialized DQN strategy, and the strategies it evolved within the running time is still highly exploitable in the large grids. This validates the effectiveness of using the more reasonable random sweeping/parameterized heuristic strategies as the initial strategies in DeDOL.  
We also note DeDOL with local mode (either local + global retraining or pure local) achieves the highest defender's expected utility in all settings. This suggests that the strategies obtained in local modes are indeed very effective and serve as good building blocks to improve the strategy quality after returning to global mode.


\presub\presec
\section{Discussions and Future Directions}
\postsub
We discuss a few questions the reader may have and some future directions.
%
First, policy gradient performs poorly in GSG-I because it learns an average of all possible sweeping routes.
Second, training DQNs is time-consuming. Though we have shown promising utility improvements, approximating NE definitely needs more iterations. Third, the global best response of an NE strategy computed in one local mode may actually be in another local mode. To address this, we hope to find a method to automatically restrict the global best response being in the current mode, which we leave for future research.
Another future direction is to consider maximum entropy Nash equilibria as the meta-strategy.
Finally, DeDOL is proposed for zero-sum GSG-I, but we expect it can be adapted to general-sum GSG-I, especially when the game is close to zero-sum.

\presec\presec
\section{Acknowledgement}
\postsec
The Azure computing resources are provided by Microsoft for Research AI for Earth award program.

\vspace{-6pt}

\small
\bibliography{aaai19}
\bibliographystyle{aaai}
\clearpage
\newpage
\normalsize

\newpage
\section{Appendices}

\subsection{A. Network Architecture of $5 \times 5$ and $3 \times 3$ Grid}
\label{appendix:network_architecture}
For the $5 \times 5$ game, the first hidden layer is a convolutional layer with 16 filters of size $3 \times 3$ and strides $1 \times 1$. The second layer is a convolutional layer with 32 filters of size $2 \times 2$ and strides $2 \times 2$. Each hidden layer is followed by a $relu$ non-linear transformation and a max-pooling layer. The output layer is a fully-connected dense layer which transforms the hidden representation of the state to the Q-value for each action.
For the $3 \times 3$ grid, the only difference is that the filter of the first convolution layer has size of $2 \times 2$ instead of $3 \times 3$, and the rest of the network architecture remains the same.

When incorporated with the dueling network structure, the second convolution layer is connected to two separate fully-connected dense layers that estimate (scalar) state-value and the (vector) advantages for each action, respectively. The output layer simply adds together the state-value and the advantages for each action to get the final Q-values.

\subsection{B. DQN Training}
\label{appendix:DQN_parameter}
Table \ref{tab:PADQN_parameter} and \ref{tab:PODQN_parameter} show the combination of parameters we have tuned on the training of Dueling double DQN. The DQN uses an epsilon-greedy policy for exploration. The exploration rate $\epsilon$ is initially set to 1 and decays by 0.05 every 15000, 15000, 5000 episodes in $7\times7$, $5\times5$ and $3 \times 3$ grids, respectively, to the final value of 0.1.
We use the Adam optimizer \cite{kingma2015adam:} to train the neural nets, with  $\beta_1=0.9$ and $\beta_2=0.999$. All gradients are clipped according to their 2-norm to a threshold of 2. For the vanilla double DQN and actor critic algorithm, we tuned its parameters in a similar way.

\begin{table*}[h]
\centering
\begin{tabular}{c|c|c|c}
Parameter & $7 \times 7$ & $5 \times 5$ & $3 \times 3$ \\ \hline 
Learning rate & \{1e-4, 5e-5, 1e-6\} $\rightarrow$ 1e-4 & \{1e-4, 5e-5, 1e-6\} $\rightarrow$ 1e-4 & \{1e-4, 5e-5, 1e-6\} $\rightarrow$ 5e-5 \\ \hline
Replay buffer size & \{1e5, 1.5e5, 2e5\} $\rightarrow$ 2e5 & 
\{2e4, 4e4, 5e4\} $\rightarrow$ 5e4 & \{5e3, 8e3, 1e4\} $\rightarrow$ 1e4 \\ \hline
Batch size &  32  & 32  & 32 \\ \hline
Target update step & \{1000, 2000\} $\rightarrow$ 1000 & \{1000, 2000\} $\rightarrow$ 1000 & \{1000, 2000\} $\rightarrow$ 1000 \\ \hline 
Episodes per iteration & 3e5 & 3e5 & 1e5 \\
\end{tabular}
\caption{The parameters of patroller dueling double DQN. Values in the curly braces are the ones we tuned, and the value behind the arrow is the one we settled on.}
\label{tab:PADQN_parameter}
\end{table*}

\begin{table*}[h]
\centering
\begin{tabular}{c|c|c|c}
Parameter & $7 \times 7$ & $5 \times 5$ & $3 \times 3$ \\ \hline 
Learning rate & \{1e-4, 5e-5, 1e-6\} $\rightarrow$ 5e-5 & \{1e-4, 5e-5, 1e-6\} $\rightarrow$ 5e-5 & \{1e-4, 5e-5, 1e-6\} $\rightarrow$ 5e-5 \\ \hline
Replay buffer size & \{1e5, 1.5e5, 2e5\} $\rightarrow$ 1e5 & 
\{2e4, 4e4, 5e4\} $\rightarrow$ 4e4 & \{5e3, 8e3, 1e4\} $\rightarrow$ 8e3 \\ \hline
Batch size &  32  & 32  & 32 \\ \hline
Target update step & \{1000, 2000\} $\rightarrow$ 1000 & \{1000, 2000\} $\rightarrow$ 1000 & \{1000, 2000\} $\rightarrow$ 1000 \\ \hline 
Episodes per iteration & 3e5 & 3e5 & 1e5 \\
\end{tabular}
\caption{The parameters of poacher dueling double DQN. Values in the curly braces are the ones we tuned, and the value behind the arrow is the one we settled on.}
\label{tab:PODQN_parameter}
\end{table*}

\subsection{C. Grid Worlds}
\label{appendix:maps}
\begin{figure*} 
\centering
\begin{tabular}{cc}
\includegraphics[scale=0.6]{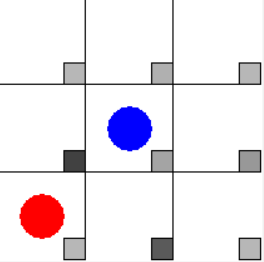} & 
\includegraphics[scale=0.6]{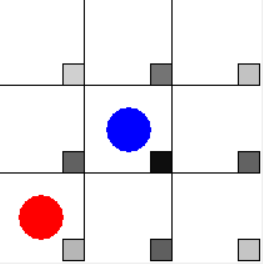} \\
& \\
(a)  3x3 grid world of random animal density. & (b) 3x3 grid world of mixture Gaussian animal density.\\
& \\
\includegraphics[scale=0.6]{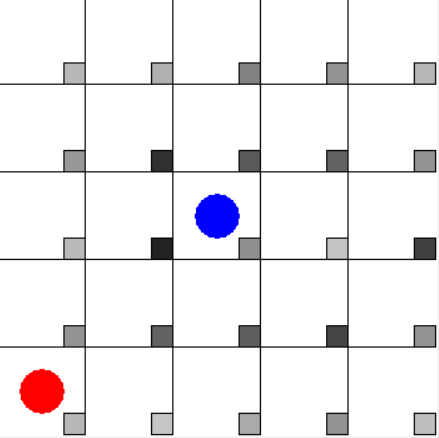} & 
\includegraphics[scale=0.6]{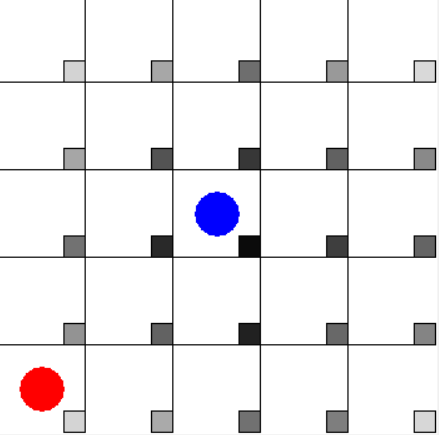}\\
& \\
(c) 5x5 grid world of random animal density.& 
(d) 5x5 grid world of mixture Gaussian animal density. \\
& \\
\includegraphics[scale=0.5]{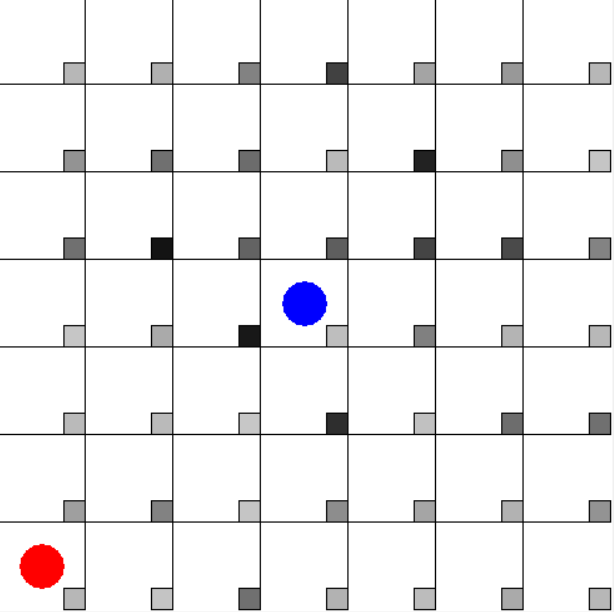}& 
\includegraphics[scale=0.5]{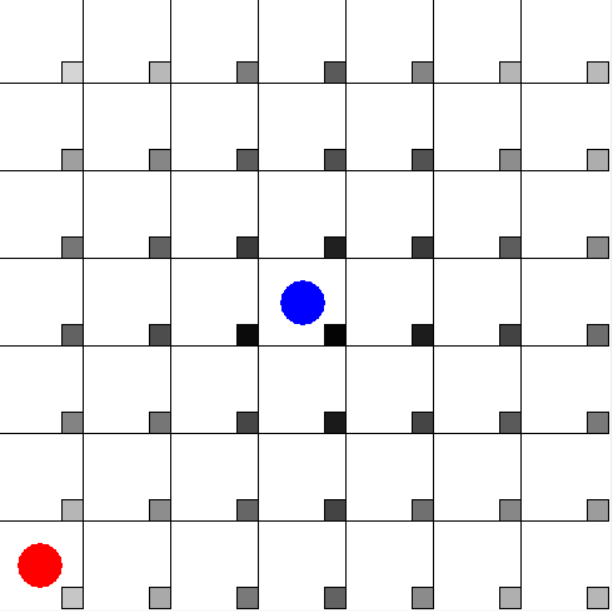}\\
& \\
(e) 7x7 grid world of random animal density.& 
(f) 7x7 grid world of mixture Gaussian animal density. \\
\end{tabular}
\caption{The illustration of grid worlds with different sizes and animal density types. Animal densities are clipped to be low around the enter point to match the real-world scenarios. For mixture Gaussian map, there are two mountain ranges, one is the horizontal middle row, and another is vertical middle column.}
\label{fig:grid_world}
\end{figure*}

Figure \ref{fig:grid_world} shows the grid worlds of different sizes and animal density types.

\subsection{D. Visualization of Poacher's Strategy}
\label{appendix:visualization}

\begin{figure*}
\centering
\begin{tabular}{cccccc}
\includegraphics[width=.3\textwidth]{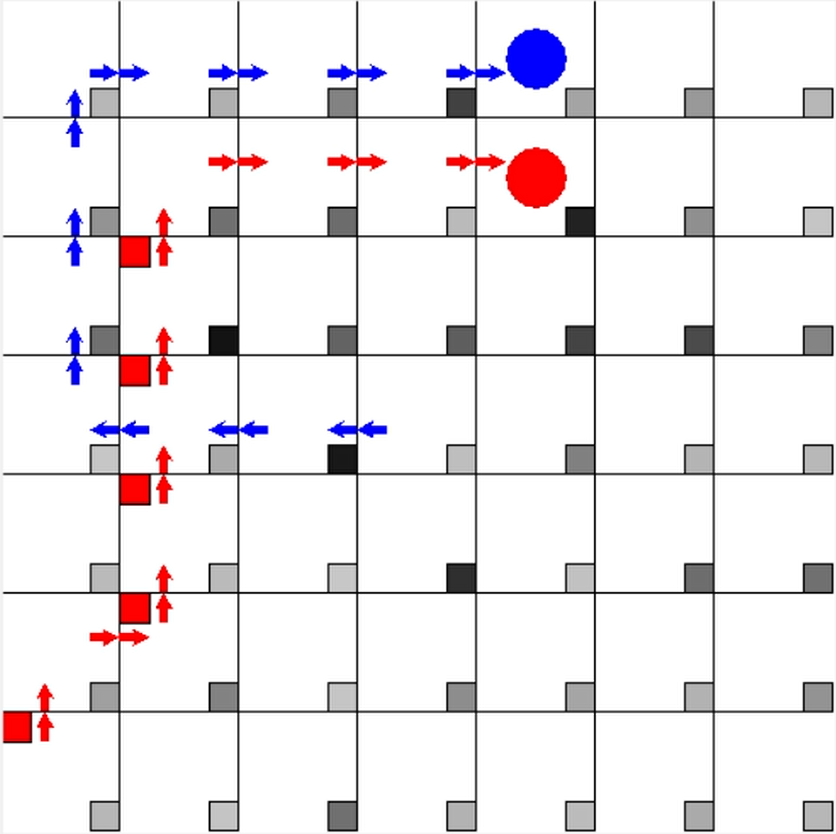}
& \quad
\includegraphics[width=.3\textwidth]{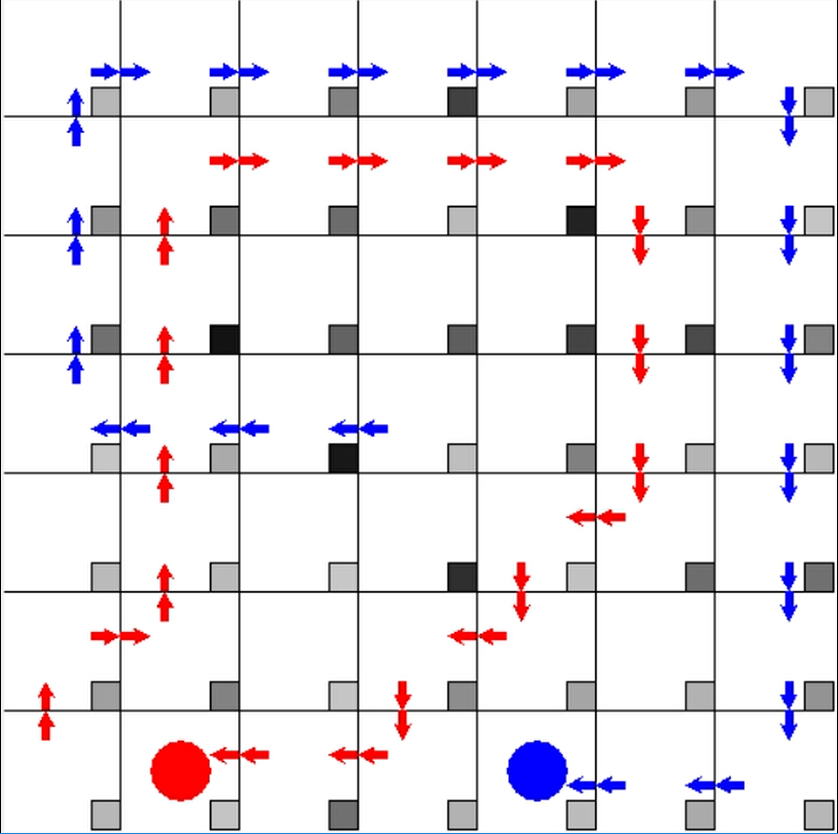}
& \quad
\includegraphics[width=.3\textwidth]{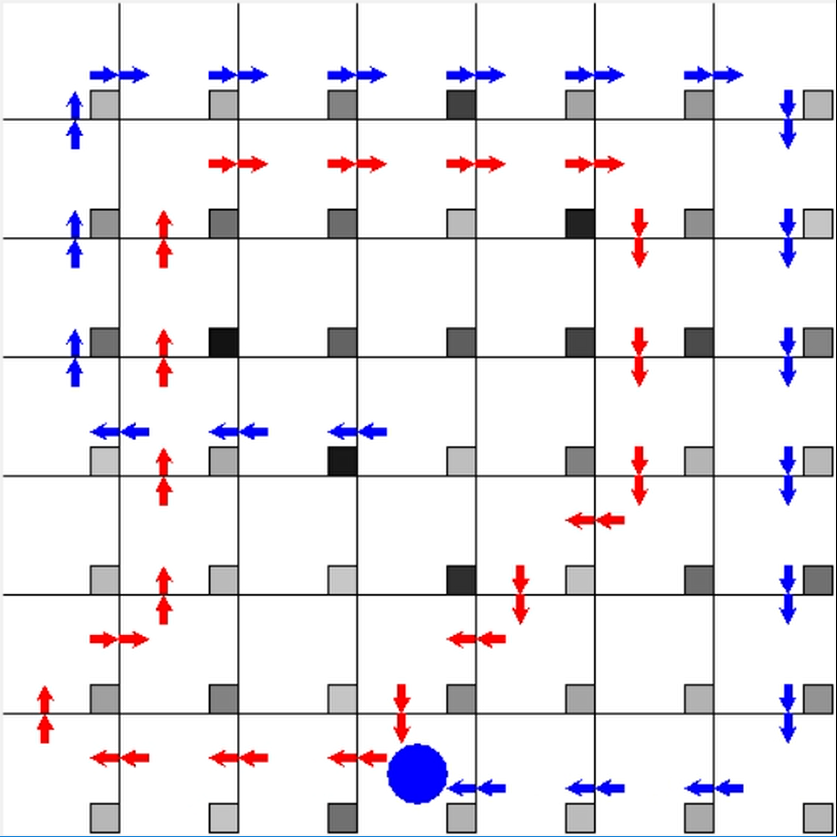}
\\
& \\
(a) At time step 9 & \quad (b) At time step 19 \quad & (c) At time step 20 \\
\\
\includegraphics[width=.3\textwidth]{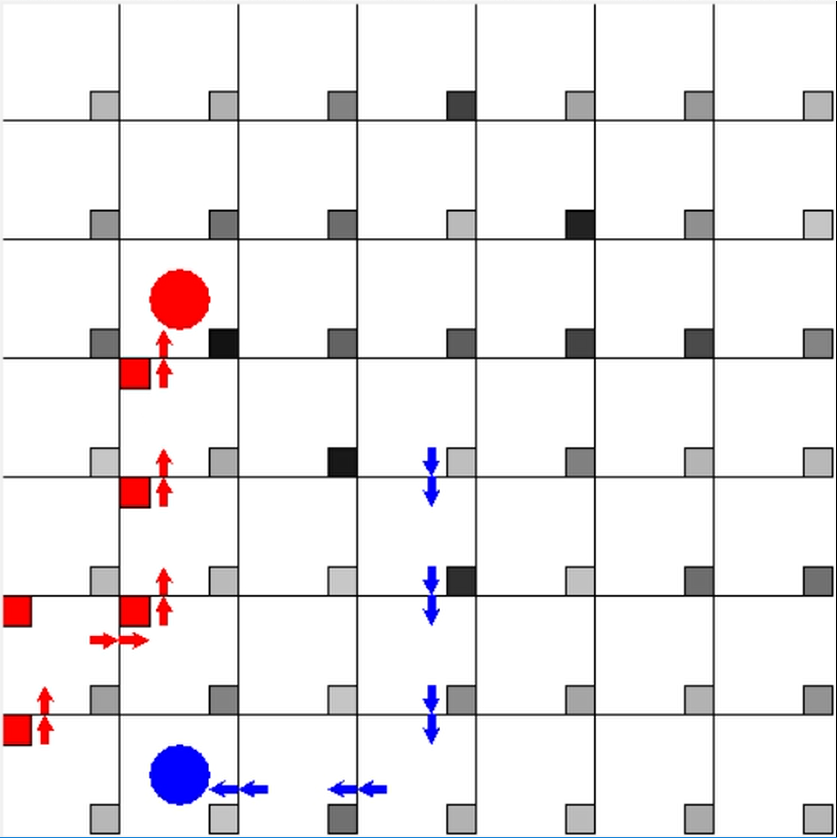}
& \quad
\includegraphics[width=.3\textwidth]{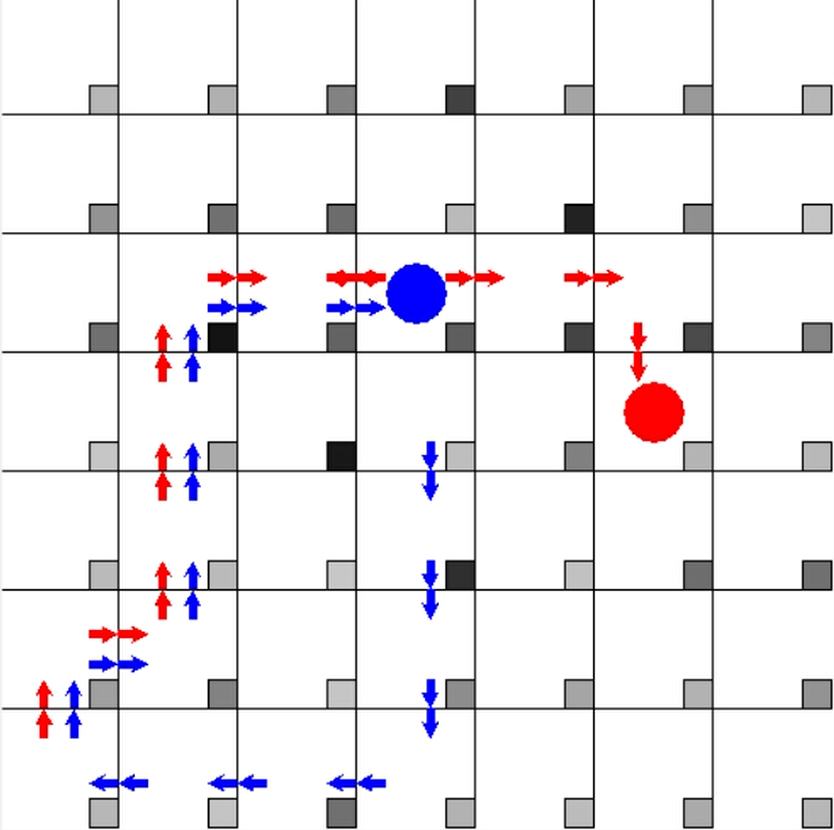}
& \quad
\includegraphics[width=.3\textwidth]{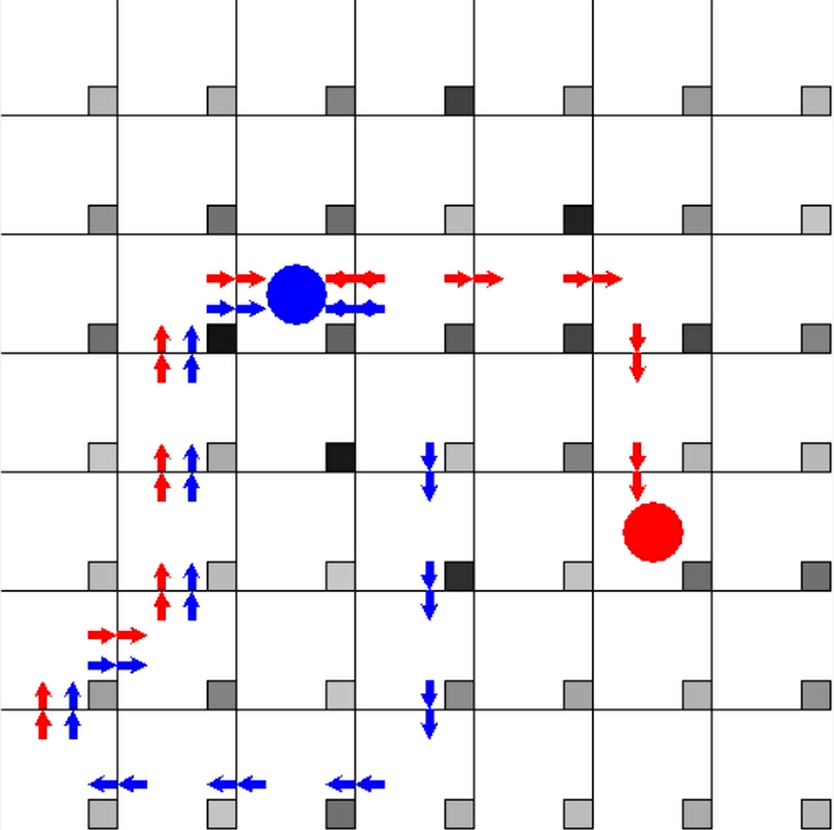}
\\
& \\
(a) At time step 5 & \quad (b) At time step 13 \quad & (c) At time step 14 \\
\end{tabular}
\caption{The learned Poacher DQN strategy against a random sweeping Patroller. 
The first row shows the case when the random sweeping patroller chooses a wrong direction to patrol. 
The second row shows the case when the random sweeping patroller chooses the right direction to patrol. 
}
\label{fig:poDQNstrategy}
\end{figure*}

Figure \ref{fig:poDQNstrategy} shows one of the learned poacher DQN strategies against a random sweeping patroller. It can be observed that the poacher DQN cleverly learns to move only in the inner part of the grid world, and leave as few as footprints on the border to avoid the chasing of the patroller. 

The first row of Figure~\ref{fig:poDQNstrategy} shows the case when the random sweeping patroller chooses a wrong direction to patrol. In such a case, the poacher judges from the footprints it observed that the patroller may have chosen a wrong direction, moves along an inner circle, places snares along the way and returns home safely. Moving along the inner circle avoids leaving footprints on the border, and thus helps it to escape the chasing of the patroller.
The second row of Figure~\ref{fig:poDQNstrategy} shows the case when the random sweeping patroller chooses the right direction to patrol. The poacher still moves along an inner circle, and it possibly judges from the fact that it observes no footprint along the way the patroller is chasing after it. Therefore, at time step 7, it turns back and leaves a fake footprint to confuse the patroller. And indeed the random sweeping patroller is confused by the turned-back footprint, which gives the poacher more time to run away. 

\subsection{E. Detailed Defender Expected Utility in DeDOL Iterations}
\label{appendix:detailed_DO}
Table~\ref{tab:detailed_DO_global5577} and~\ref{tab:detailed_DO_global33} show the defender's expected utility at each iteration in the pure global mode version of the DeDOL algorithm. Table~\ref{tab:detailed_DO_separate+global5577} shows the results in the local + global mode version of DeDOL. The number of iterations refers in the table to that completed in the global mode.

\begin{table}
\small
\centering
\begin{tabular}{c|c|c|c|c|c}
Iteration & 0 & 1 & 2 & 3 &  4 \\ \hline
$5 \times 5$ Random & -1.912 & \textbf{-1.630} & -3.507 & -4.800 & -1.652  \\ \hline
$5 \times 5$ Gaussian &-1.614 & -3.037 & -2.355 & \textbf{-0.434} &  -0.992 \\ \hline
$7 \times 7$ Random & -4.06 & -6.416 & \textbf{-2.003} & -3.465 & -5.134\\ \hline
$7 \times 7$ Gaussian & -4.250 & -7.761 & \textbf{-4.154} & -9.144 & -4.185 \\ 
\end{tabular}
\caption{The defender's expected utility at each iteration of DeDOL, pure global mode, on $5 \times 5$ and $7 \times 7$ grid. }
\label{tab:detailed_DO_global5577}
\end{table}

\begin{table}
\centering
\scriptsize
\begin{tabular}{c|c|c|c|c|c|c}
Iteration & 0 & 1 & 2 & 3 &  4 & 5 \\ \hline
$3 \times 3$ Random & -0.035 & -1.231 & 0.532 & 0.113 & 0.191 & 0.728\\ \hline
$3 \times 3$ Gaussian & -0.092 & -0.175 & -0.121 & 0.694 &  \textbf{0.745} &  0.732 \\ \hline
Iteration & 6 & 7 & 8 & 9 &  10 & 11 \\ \hline
$3 \times 3$ Random & 0.712 & 0.181 & 0.179 & \textbf{0.734} & 0.023 & 0.226 \\ \hline
$3 \times 3$ Gaussian & 0.011 & 0.472 & 0.435 & 0.253 & 0.141 & 0.133 \\ \hline
Iteration & 12 & 13 & 14 & 15 &  16  \\ \hline
$3 \times 3$ Random & 0.542 & 0.493 & 0.648 & 0.361 & 0.612 &  \\ \hline
$3 \times 3$ Gaussian & 0.141 & 0.246 & 0.521 & 0.642 & 0.711  \\
\end{tabular}
\caption{The defender's expected utility at each iteration of DeDOL, pure global mode, on $3 \times 3$ grid. }
\label{tab:detailed_DO_global33}
\end{table}

\begin{table}
\centering
\begin{tabular}{c|c|c|c}
Iteration (of global mode) & 0 & 1  & 2 \\ \hline
$3 \times 3$ Random & 0.332 & \textbf{0.851} & 0.694 \\ \hline
$3 \times 3$ Gaussian & 0.173 & 0.422 & \textbf{0.862}  \\ \hline
$5 \times 5$ Random & \textbf{-0.423} & -2.177 & - \\ \hline
$5 \times 5$ Gaussian & \textbf{0.604} & -0.978 & -\\ \hline
$7 \times 7$ Random & -2.046 & \textbf{-0.543}  & -\\ \hline
$7 \times 7$ Gaussian & \textbf{-2.353} & -4.065  & -\\ 
\end{tabular}
\caption{The defender's expected utility at each DeDOL iteration, local + global mode, on $3 \times 3$, $5 \times 5$ and $7 \times 7$ grid. }
\label{tab:detailed_DO_separate+global5577}
\end{table}

\subsection{F. Computing Exact Best Response in $3 \times 3$ Grid} 
\label{appendix:33bestresponse}
In the $3\times 3$ grid world with 4 time steps, the game tree has roughly $4.5\times 10^7$ nodes. We can compute an exact best response given a patroller strategy here, following the methods proposed in~\cite{bosansky2013double-oracle}. Below is a brief summary of the method, and the reader may refer to the original paper for full details. 

We refer to the player that is computing a best response as the searching player. To compute the exact best response for him,
we first build the full game tree for our GSG-I game. Then, we use depth-first search to traverse the game tree and recursively run the ``Evaluate'' function which returns the state value, i.e., the searching player's best response expected utility when reaching that node, and also returns the optimal action at that state if it belongs to the searching player. 

More concretely, if node $h$ is a chance node, then the ``Evaluate'' function recursively evaluates the succeeding nodes, computes the expected value of node $h$ as a sum of the values of the successors weighted by the fixed probability distribution associated with node $h$, and propagates this expected value to the predecessor. If node $h$ is assigned to the opponent, the algorithm acts similarly, but the probability distribution is given by the strategy of the opponent.

If the function evaluates node $h$ associated with the searching player, the algorithm first recursively evaluates the succeeding nodes of all the nodes in the information set where this node $h$ belongs. Then it selects the best action for the information set since the searching player does not know exactly which state or node he is in. So he needs to check the expected utility of each action. The expected utility of an action a is the sum of the probability of being in each specific node $h'$ multiplied by the expected value of the successor node of $h'$ when action a is taken. Once the best action $a'$ is found, the expected value of node $h$ is simply the value of the successor node of h when the optimal action $a'$ is taken.
The probability of being in a specific node $h'$ in this information set is given by the probability of the chance player and the opponent's strategy (i.e., multiplying all the behavior strategy probabilities along the path).


\end{document}